\documentstyle[12pt]{article}
\begin{document}
\setlength{\topmargin}{1.0cm}
 \newcommand\beq{\begin{equation}}
  \newcommand\eeq{\end{equation}}
 \newcommand\beqn{\begin{eqnarray}}
  \newcommand\eeqn{\end{eqnarray}}
 \newcommand{\doublespace} {
  \renewcommand{\baselinestretch} {1.6} \large\normalsize}
  \renewcommand{\thefootnote}{\fnsymbol{footnote}}
  \def\sel{\sigma_{el}^{VN}}
 \def\sin{\sigma_{in}^{VN}}
  \def\stot{\sigma_{tot}^{VN}}
 \def\inf{\int_{-\infty}^{\infty}}

 \setlength{\baselineskip}{0.70cm}

\begin{center}

 {\huge{\bf The Role of the Coherence Length\\

 \smallskip

 in Exclusive  Electroproduction\\

 \smallskip

 of Vector Mesons off Nuclei\\

  \bigskip

 at Moderate Energies}\footnote{Invited talk given by B.~Kopeliovich
at the ELFE (Electron Lab. For Europe) Workshop,\\
Cambridge, 22-29 July, 1995}} \\

\vspace{1cm}

{\large J\"org~H\"ufner}\\
\medskip

{\it Instut f\"ur Theoretische Physik der Universit\"at ,\\
Philosophenweg 19, 69120 Heidelberg, Germany}\\
\bigskip

 {\large Boris~Kopeliovich}\footnote{On leave from
 Joint
Institute for Nuclear Research, Dubna, 141980
 Moscow Region, Russia.\\
E-mail: bzk@dxnhd1.mpi-hd.mpg.de}\\
 \medskip

 {\it Max-Planck
Institut f\"ur Kernphysik,\\ Postfach 103980,
 69029 Heidelberg,
Germany}\\
 \bigskip

 {\large Jan Nemchik\footnote{On leave from
Institute of Experimental Physics SAV,
 Solovjevova 47, CS-04353 Kosice,
Slovakia}}\\
 \medskip

 {\it Dipartimento di Fisica Teorica,
Universit\`a di Torino\\
 and INFN, Sezione di Torino, I-10125, Torino,
Italy}
 \\

\vspace{1.0cm}

 \large {\bf ABSTRACT}

\end{center}

The variation of the coherence length $l_c=2\nu/(Q^2+m_V^2)$
in virtual photoproduction of vector mesons
off  nuclei as a function of the photon energy $\nu$ or
virtuality $Q^2$ results in dramatic changes in the values of
the nuclear transparency. Color transparency effects can be easily
mixed up with the effects of the coherence length
in incoherent photoproduction on nuclei.


  \vfill

\setlength{\textheight}{25.0cm}

\pagebreak

 \setlength{\topmargin}{-1.5cm}

\section{Color transparency in exclusive
electroproduction of vector mesons off nuclei}

Virtual photoproduction of vector mesons off nuclei
 was suggested in \cite{knnz} as an
 effective way for detecting color transparency
 (CT)
effects.  One can change the size of the
 produced wave packet by varying $Q^2$,
but keeping
 the photon energy high and fixed.  This is
 different from
quasielastic $(e,e'p)$ and $(p,2p)$
 reactions, where one may have high energy
only at the
 expense of high $Q^2$, i.e.  very small cross
 section.  Recently
the E665 experiment \cite{e665} has
 claimed to confirm, although with a poor
statistics, the CT effects predicted in
 \cite{knnz}.

 Analogous
experiments are planned in the
 CEBAF-HERMES-ELFE energy range: Rising
$Q^2$-dependence of nuclear transparency is
 expected to be a signal of
CT. However, we would
 like to warn in this paper against straightforward
application of the high-energy predictions of
 \cite{knnz} at moderate
energies.  We demonstrate
 that also effects of the
coherence length provide a steep
variation of nuclear transparency with $Q^2$,
 which can mock CT effects or
make interpretation
 of the data more ambiguous.

 We work in the Glauber
approximation, which is
 usually supposed to be a base line for CT studies.
The correct formula for incoherent (virtual)
 photoproduction of vector mesons
off nuclei was
 not known in the literature, and we present it
 here.

\section{Coherence length}

 Vector mesons produced at different longitudinal
 coordinates have relative
phase shifts $q_L\Delta z$ due
to the difference in the virtual photon
 and the meson longitudinal momenta
 $q_L=(Q^2+m_V^2)/2\nu$.  Only those mesons
interfere constructively which are produced
 sufficiently close to each other:
$\Delta z\leq
 l_c$, where
 \beq l_c={1\over q_L}=\frac{2\nu}{Q^2+m_V^2}
\label{4} \eeq is called coherence length.
It can be also interpreted as a lifetime of the
 hadronic fluctuation of the
photon.  If $l_c$ is
 much shorter
 that the mean inter-nucleon distance,
there is no
 nuclear shadowing in the initial state.
 However, if $l_c$ is
much longer than the mean
 free path of the vector meson in the nucleus
or the nuclear radius,
nuclear suppression is expected to be stronger
 than that in
the low-energy
 limit.

 Thus, the energy- and
 $Q^2$-variation of $l_c$
results in dramatic
 changes in the nuclear transparency.
These variations may easily be mixed
 up with CT effects in some cases.
For the photoproduction of charmonium
the transition
 region covers the energies from a few tens to a few
 hundreds of GeV. For light
 vector meson ($\rho, \omega,\phi$)
the transition energy range is an order of magnitude lower.

\section{Incoherent electroproduction of vector
mesons}

 The formula for the exclusive incoherent
 production of vector mesons which
incorporates the
 effects of the coherence length is derived for the
 first
time\footnote{One can find a formula for
 incoherent
 photoproduction of
vector mesons in ref.
 \cite{bauer} which differs from our eqs.
(\ref{6})--(\ref{8}), and we consider it as
 incorrect.  That formula
underestimates available
 data on real photoproduction of $\rho$ off nuclei
(see corresponding discussion in ref.
 \cite{bauer}). Our calculations nicely
agree with
 the data.} in ref.  \cite{hkn}.  The nuclear
 transparency for
the cross section integrated over
 momentum transfer can be represented as,

 \beq Tr_{inc}(\gamma^*A\ss VX)=Tr_1+Tr_2-Tr_{coh}\
 , \label{6} \eeq where
the first term

 \beq Tr_1=\frac{1}{\sin} \int d^2b \left
 [1-e^{-\sin
T(b)}\right ] \label{7} \eeq
 corresponds to the case where
the vector meson is produced on the
 same nucleon in both
interfering
 amplitudes.

If the nucleons are different,
 the corresponding term $Tr_2$ in eq.
(\ref{6})
 reads,

 \beqn & & Tr_2 = \frac{\stot}{2\sel}(\sin-\sel)
 \int
d^2b\ \inf dz_2\ \rho(b,z_2)
 \int_{-\infty}^{z_2} dz_1\ \rho(b,z_1)\ \times
\nonumber\\ & & e^{iq_L(z_2-z_1)}\ \exp\left[
 -{1\over 2}\stot\
\int_{z_1}^{z_2}dz
 \rho(b,z)\right] \exp\left[ -\sin\
 \int_{z_2}^{\infty}
dz\ \rho(b,z)\right] \label{8}
 \eeqn

 The two first terms in eq. (\ref{6})
correspond to
 the sum over all final states of the nucleus
 via completeness.  For this
 reason the third term which
represents coherent production
has to be subtracted in eq.  (\ref{6}).

 \beq
Tr^{coh}(\gamma^*A\ss VA)=\frac{(\stot)^2}{4\sel}
 \inf d^2b\left |\int dz\
\rho(b,z)\ e^{iq_Lz}
 \exp\left[-{1\over 2}\stot\int_z^{\infty}
dz'\rho(b,z')\right ]\right |^2
\label{5}
\eeq

 Here $\rho(b,z)$ is the nuclear density and
 $b$ the impact
parameter. $T(b)=\inf
 dz\rho(b,z)$ is the nuclear thickness function.

 In
low-energy limit only the term eq.
 (\ref{7}) survives. In the
high-energy limit, $q_L
 \ll 1/R_A$, the nuclear transparency is the same as
in quasielastic V-A scattering, what can
 be interpreted as a result
of the
fluctuation $\gamma^* \ss V$
 long in advance of the
 nucleus.
Thus nuclear transparency of incoherent
electroproduction of vector mesons
is expected to decrease with $\nu$. This was predicted
in \cite{kz,bkmnz} for the energy dependence of $J/\Psi$
photoproduction and confirmed by the NMC experiment as is shown
in fig. 1 \cite{bkmnz}.
Another example of the energy dependence of the nuclear transparency
for real photoproduction of $J/\Psi$ on Beryllium is
shown in fig. 2.

\begin{figure}[thb]
\includegraphics{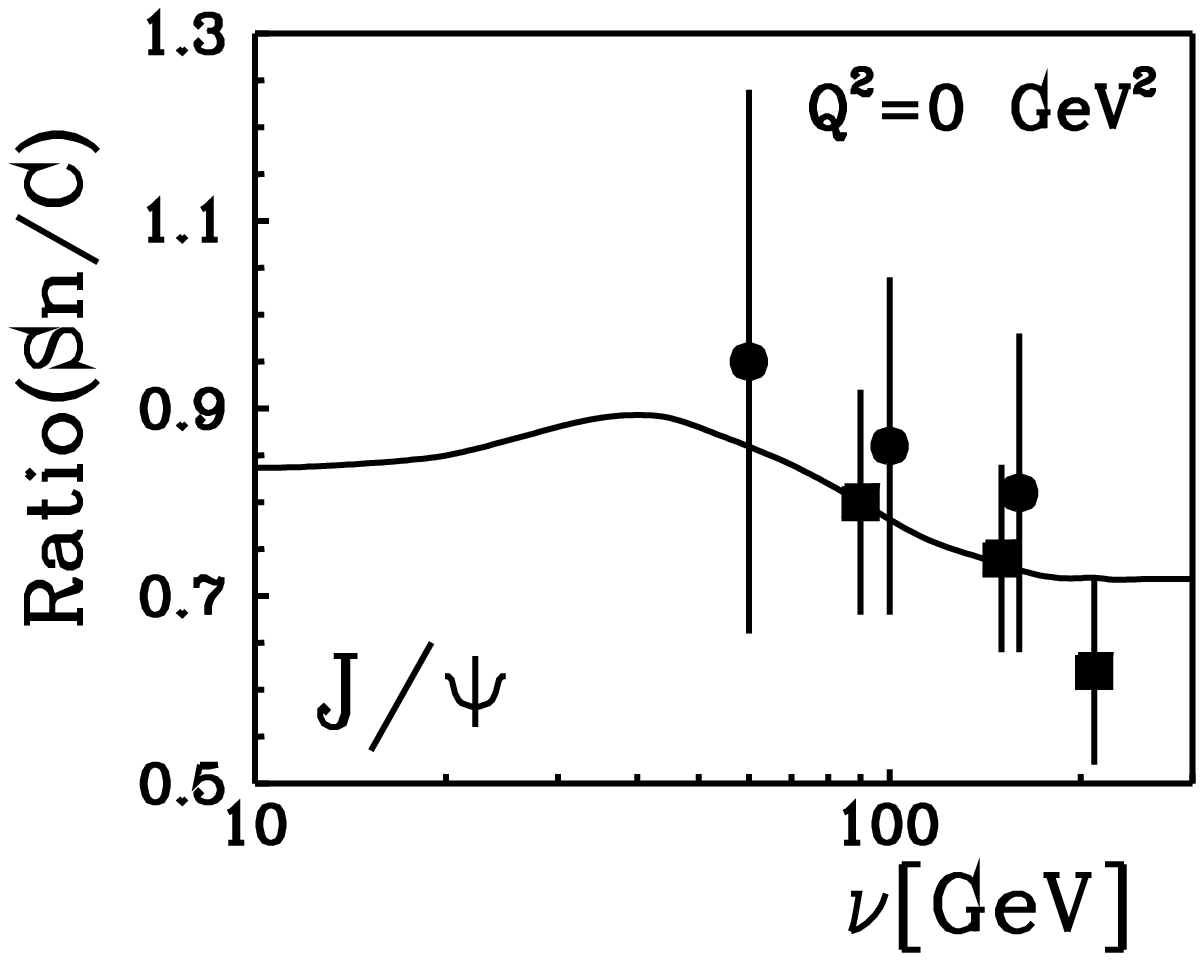}
\includegraphics{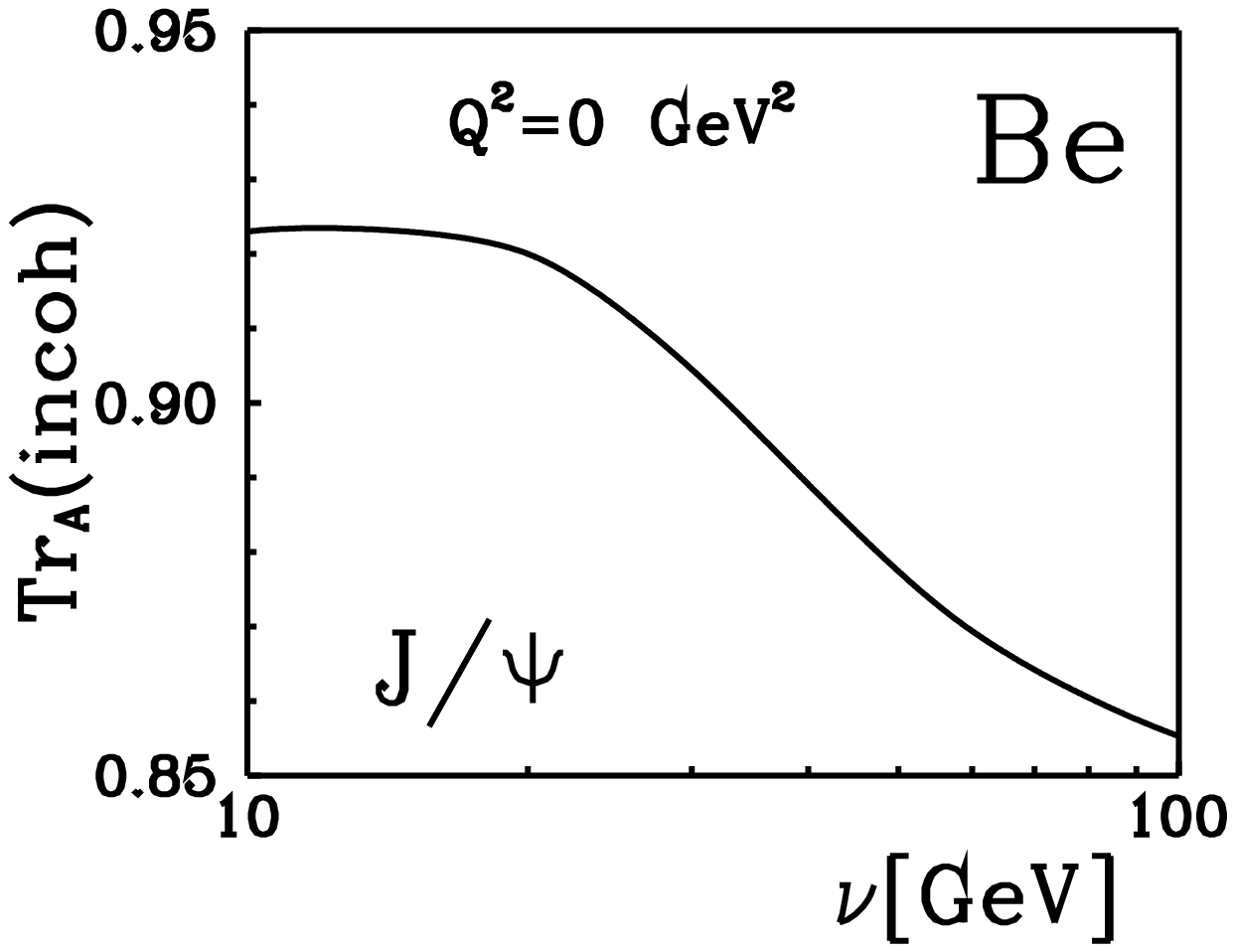}
\begin{center}
\vspace{6.5cm}
\parbox{13cm}
{\caption{Energy-dependence of ratio
of $J/\Psi$
real photoproduction  cross sections on $Sn$ to $C$
[7].
The non-trivial dependence only  originates
from the variation of the coherence length with $\nu$}
\caption{The same as in fig. 1 for nuclear
transparency on Be.}}
\end{center}
\end{figure}

\begin{figure}[thb]
\includegraphics{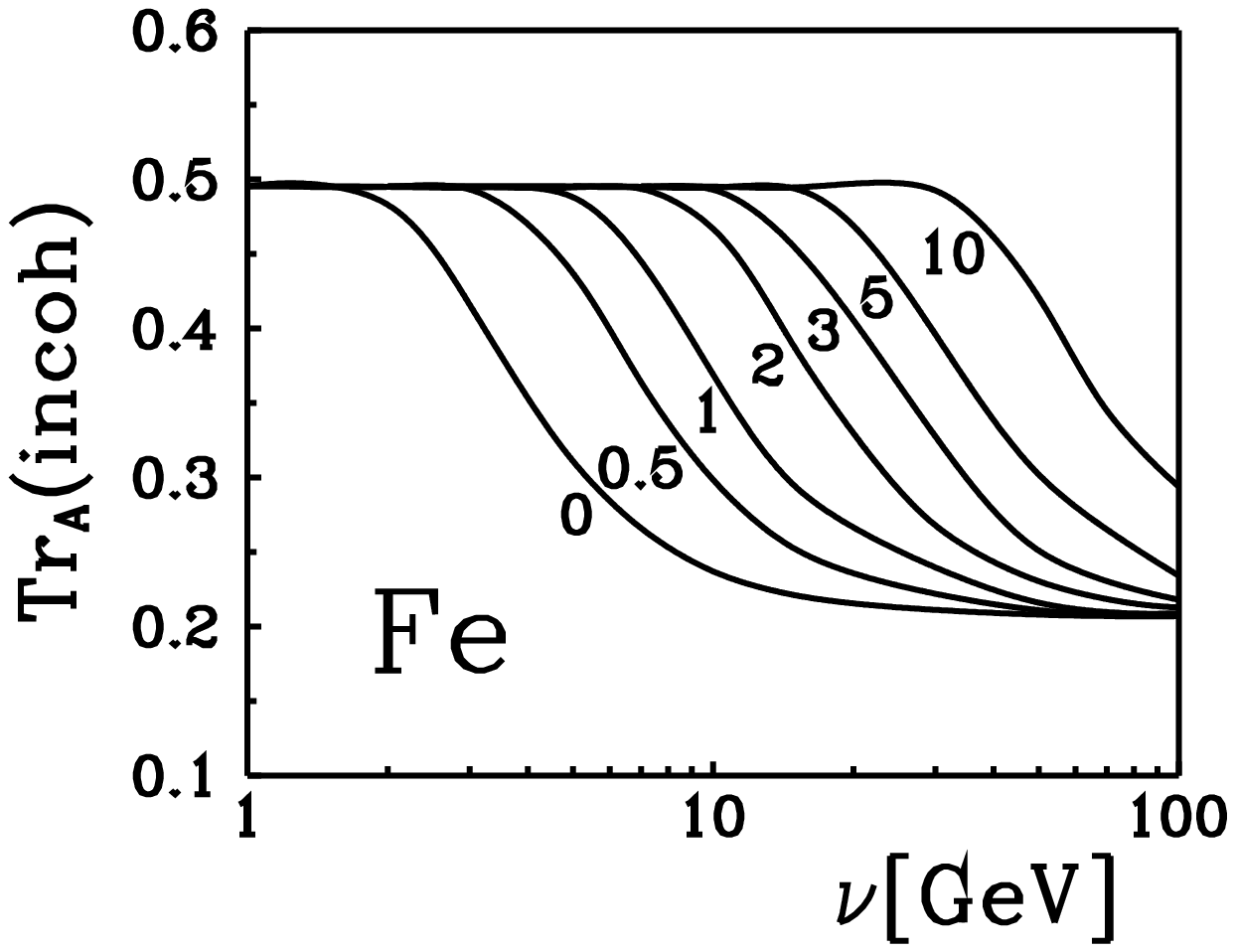}
\includegraphics{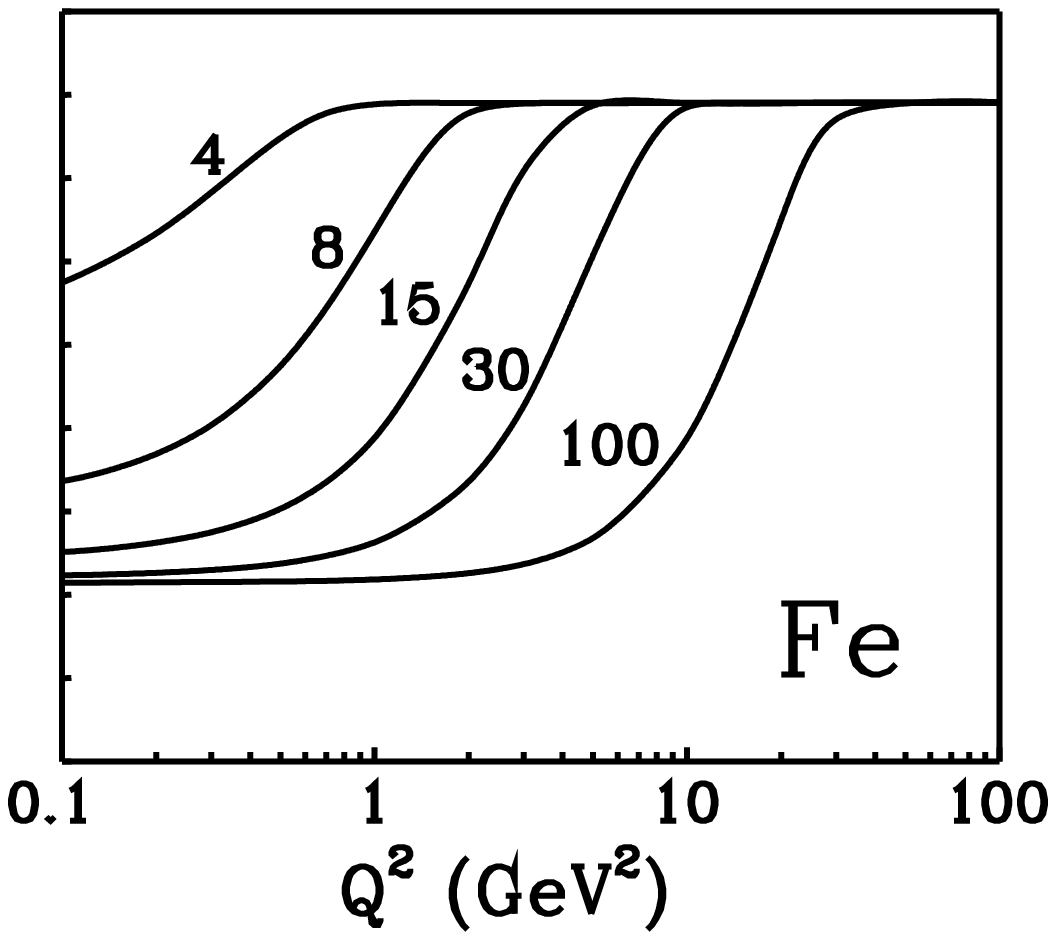}
\begin{center}
\vspace{6.5cm}
\parbox{13cm}
{\caption{Energy dependence of nuclear transparency
in virtual incoherent
photoproduction of $\rho$-meson on $Fe$
for different values of $Q^2$.}
\caption{$Q^2$-dependence of nuclear transparency
in virtual incoherent
photoproduction of $\rho$-meson on $Fe$
for different values of $\nu$}}
\end{center}
\end{figure}

Analogous calculations for the electroproduction of $\rho$-mesons
on Iron are shown in fig. 3as function of energy at different
values of $Q^2$. The energy variation of
nuclear transparency is much steeper than for charmonium
due to the larger cross section.

The shrinkage of the coherence length
 as a function of $Q^2$ at
fixed energy, causes a growth
of nuclear transparency with  $Q^2$,
a behaviour which resembles CT and
should be taken into account searching for CT.
 Examples of
$Q^2$-dependence for
 $\rho$-meson photoproduction versus
photon energy $\nu$ are shown in fig. 4
Nuclear transparency steeply increases and
 then saturates at high $Q^2$.
 Note that nuclear transparency
cannot reach unity at medium energies, even
 in presence of CT.  $Tr(Q^2)$
saturates at about
 the same level as is shown in fig. 4.
Indeed, the predicted
 saturation in the
Glauber approximation
signals that the coherence length
 becomes negligibly short, i.e. there
 is no room for the full CT.

Note that our prediction of steep $Q^2$ dependence of
nuclear transparency at moderate energy is in variance with
the results of \cite{bznf} which expect nearly
$Q^2$-independent nuclear transparency for electroproduction of
$\phi$-meson at $\nu = 8\ GeV$. The reason of so large discrepancy
is neglection in \cite{bznf} the effects
of coherence length, which was fixed at zero.

\section{Coherent electroproduction of vector mesons}

 In this case the nuclear transparency
has the form of eq. (\ref{5}) (see for instance \cite{bauer}),

\begin{figure}[thb]
\includegraphics{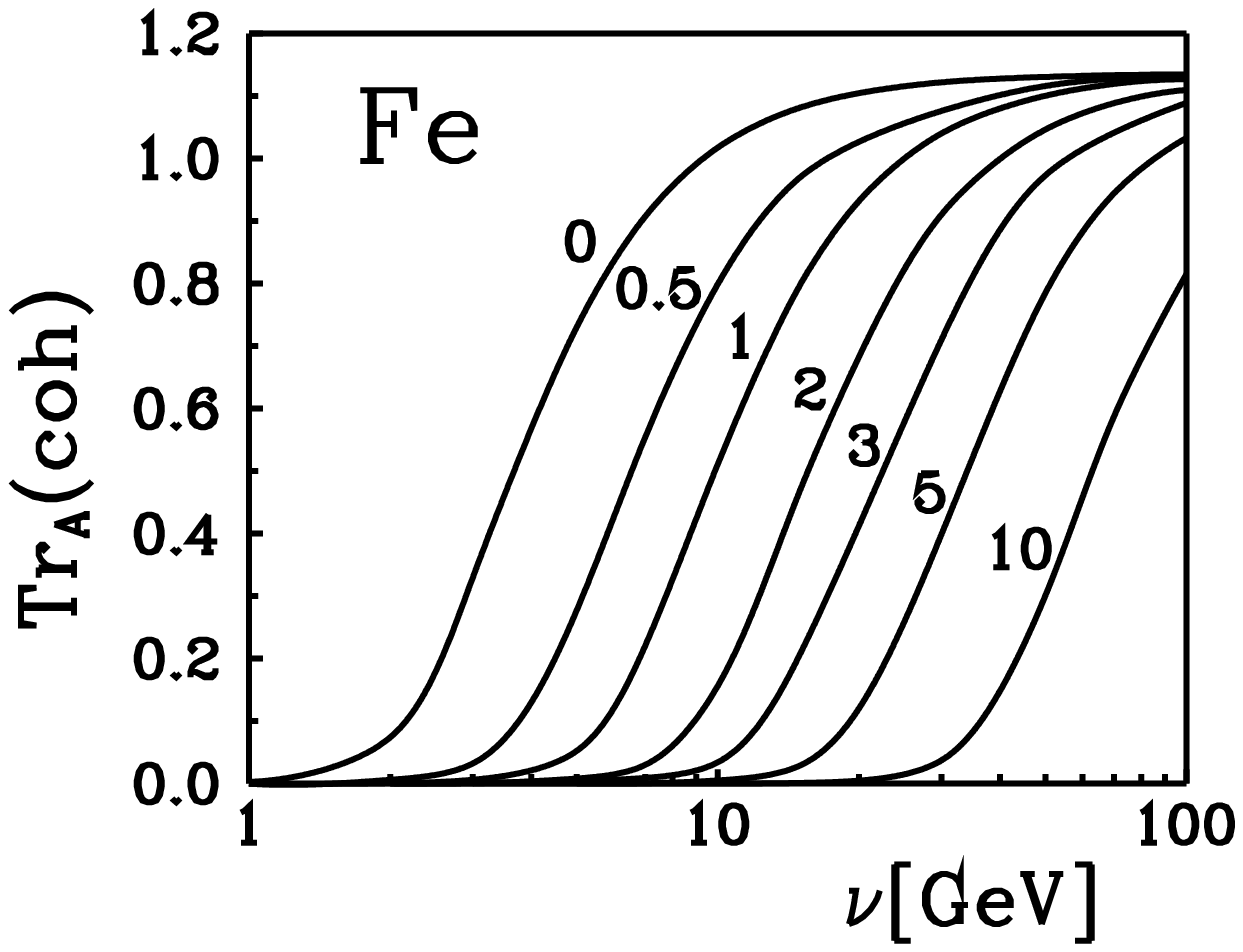}
\includegraphics{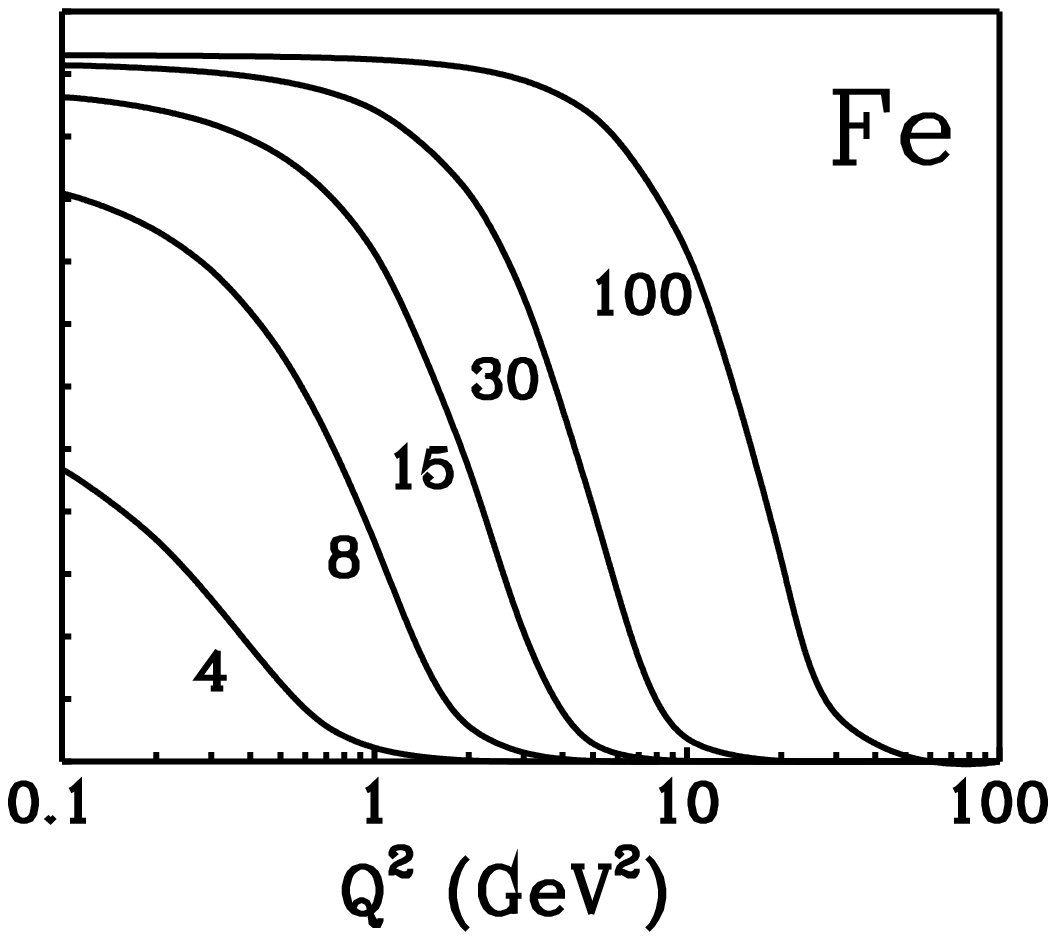}
\begin{center}
\vspace{6.5cm}
\parbox{13cm}
{\caption{The same as in fig. 3, but for coherent production.}
\caption{The same as in fig. 4, but for coherent production.}}
\end{center}
\end{figure}

 The results of
calculation of energy dependence of nuclear transparency
at different values of $Q^2$ for
coherent photoproduction of $\rho$-mesons
are plotted in fig. 5.

 We present in fig. 6 the
 Glauber model predictions for
variation of nuclear transparency as function of $Q^2$ in
 $\rho$-meson photoproduction at different
photon energies. In this case the nuclear transparency
is a decreasing function of $Q^2$, i.e. has an opposite trend
compared to what is supposed to be a signal of CT.
We conclude that coherent virtual photoproduction
of vector mesons at moderate energies
is a better reaction for CT studies that incoherent one.

\medskip

Summarising, we demonstrate the importance of coherence length effects
for incoherent and coherent exclusive electroproduction
of vector mesons off nuclei. These effects provide a steep variation
of nuclear transparency as function of energy and virtuality
of the photon.

The Glauber formula, which we use for incoherent
production is new.

We use the eikonal Glauber approximation {\it on purpose}.
Inclusion of the excited intermediate
states would be equivalent to the
CT effect. However, we are interested in
a base line for study of CT. Note that the effect
of $Q^2$-dependence of nuclear transparency in $(e,e'p)$
reaction due to inelastic shadowing, discussed in section 2,
is irrelevant in this case. Indeed, in $(e,e'p)$
it originates from energy dependence, but the photon energy
is fixed in the photoproduction experiments.

It was found in \cite{kz,knnz} that for real
photoproduction of
ground states of vector mesons these corrections
are quite small. At moderate energies they are small
even at high $Q^2$, since as is argued above the nuclear
transparency does not reach unity at high $Q^2$, but
saturates at about the same level as in the
Glauber approximation.

\medskip

{\bf Acknowledgements}: B.K. is grateful to S.~Bass
for invitation to the Workshop and a partial support. A support from
Max-Planck-Institut f\"ur Kernphysik is also acknowledged.

\end{document}